\documentclass[american,english]{elsarticle}
\usepackage[T1]{fontenc}
\usepackage[latin9]{inputenc}
\usepackage{color}
\usepackage{array}
\usepackage{multirow}
\usepackage{amsbsy}
\usepackage{amstext}
\usepackage{amssymb}
\usepackage{graphicx}

\makeatletter

\providecommand{\tabularnewline}{\\}

\makeatother

\usepackage{babel}
\begin{document}

\title{A Fortran 90 Hartree-Fock program for one-dimensional periodic $\pi$-conjugated
systems using Pariser-Parr-Pople model}

\author{Kondayya Gundra\fnref{fn1}}

\ead{naiduk@barc.gov.in}

\author{Alok Shukla}

\ead{shukla@phy.iitb.ac.in}

\fntext[fn1]{Permanent address: Theoretical Physics Division, Bhabha Atomic Research
Centre, Mumbai 400085, INDIA}

\address{Department of Physics, Indian Institute of Technology, Bombay, Powai,
Mumbai 400076, INDIA}
\begin{abstract}
Pariser-Parr-Pople (P-P-P) model Hamiltonian is employed frequently
to study the electronic structure and optical properties of $\pi$-conjugated
systems. In this paper we describe a Fortran 90 computer program which
uses the P-P-P model Hamiltonian to solve the Hartree-Fock (HF) equation
for infinitely long, one-dimensional, periodic, $\pi$-electron systems.
The code is capable of computing the band structure, as also the linear
optical absorption spectrum, by using the tight-binding (TB) and the
HF methods. Furthermore, using our program the user can solve the
HF equation in the presence of a finite external electric field, thereby,
allowing the simulation of gated systems. We apply our code to compute
various properties of polymers such as $trans$-polyacetylene ($t$-PA),
poly-\emph{para}-phenylene (PPP), and armchair and zigzag graphene
nanoribbons, in the infinite length limit. \end{abstract}
\maketitle
\textbf{Program Summary} \\
 \emph{Title of program:} ppp\_bulk.x \\
 \emph{Catalogue Identifier:} \\
 \emph{Program summary URL:} E\_F\\
 \emph{Program obtainable from:} CPC Program Library, Queen's University
of Belfast, N. Ireland \\
 \emph{Distribution format:} tar.gz\\
 \emph{Computers :} PC's/Linux\\
\emph{Linux Distribution:} Code was developed and tested on various
recent versions of 64-bit Fedora including Fedora 14 (kernel version
2.6.35.12-90)\\
 \emph{Programming language used:} Fortran 90\\
\emph{Compilers used:} Program has been tested with Intel Fortran
Compiler (non-commercial version 11.0.074) and gfortran compiler (gcc
version 4.5.1) with optimization option -O. \emph{}\\
\emph{Libraries needed: }This program needs to link with LAPACK/BLAS
libraries compiled with the same compiler as the program. For the
Intel Fortran Compiler\emph{ }we used the ACML library version 4.4.0,
while for the gfortran compiler we used the libraries supplied with
the Fedora distribution.\\
\emph{Number of bytes in distributed program, including test data,
etc.:} ... size of the tar file \\
 \emph{Number of lines in distributed program, including test data,
etc.:} ... lines in the tar file \\
 \emph{Card punching code:} ASCII\\
 \emph{Nature of physical problem:} The electronic structure of one-dimensional
periodic $\pi$-conjugated systems is an intense area of research
at present because of the tremendous interest in the physics of conjugated
polymers and graphene nanoribbons. The computer program described
in this paper provides an efficient way of solving the Hartree-Fock
equations for such systems within the P-P-P model. In addition to
the Bloch orbitals, band structure, and the density of states, the
program can also compute quantities such as the linear absorption
spectrum, and the electro-absorption spectrum of these systems. \emph{}\\
\emph{Method of Solution:} For a one-dimensional periodic $\pi$-conjugated
system lying in the $xy$-plane, the single-particle Bloch orbitals
are expressed as linear combinations of $p_{z}$-orbitals of individual
atoms. Then using various parameters defining the P-P-P Hamiltonian,
the Hartree-Fock equations are set up as a matrix eigenvalue problem
in the $k$-space. Thereby, its solutions are obtained in a self-consistent
manner, using the iterative diagonalizing technique at several $k$-points.
The band structure and the corresponding Bloch orbitals thus obtained
are used to perform a variety of calculations such as the density
of states, linear optical absorption spectrum, electro-absorption
spectrum, \textit{etc}.\\
\emph{Running Time:} Most of the examples provided take only a few
seconds to run. For a large system, however, depending on the system
size, the run time may be a few minutes to a few hours.\emph{}\\
\emph{Unusual features of the program:} None

\section{Introduction}

Conjugated molecules and polymers have been actively investigated
theoretically as well as experimentally in recent years,\citet{salem-book,sumit-review,barford-book}
because of their potential applications in manufacture of optoelectronic
devices, and solar cells\citet{polymer review}. The low-lying excitations
in such materials are characterized by the $\pi$ electrons, which
have itinerant nature, and form the energy levels near the chemical
potential (Fermi level). Recently, the field of $\pi$-conjugated
systems has received a tremendous boost with the synthesis of graphene\citet{geim-novoselov},
and its heterostructures such as graphene nanoribbons\citet{gnr-review},
which exhibit exotic transport and electronic properties, leading
to the possibility of future electronic devices based upon graphene
rather than silicon\citet{geim-nature,geim-rmp,peres-rmp}. Because
of these recent advances, theoretical studies of $\pi$-electron systems
have come to the forefront of physics. Most of the theoretical methods
used for describing the electronic structure of these materials can
be classified as: (a) fully \emph{ab initio }approaches based upon
the mean-field methods such as the density-functional theory (DFT)\citet{Son,Scuseria,yang,yang2,yang3,Prezzi}
or the Hartree-Fock (HF) method\citet{PBC-polymers,crystal,wannier,wann-pol},
and (b) approaches based upon effective $\pi$-electron models such
as the tight-binding (TB) model\citet{Nakada,fujita,Ezawa}, and the
Hubbard\citet{fernandez-hubbard,yazyev-hubbard,jung-hubbard} or the
extended Hubbard model\citet{ext-hub}. The \emph{ab initio} methods
are generally computationally intensive because they make no distinction
between the $\sigma$ and the $\pi$ electrons of the system, and
therefore require the use of large basis sets to provide a reasonable
description of the electronic structure of such systems. In case of
graphene nanoribbons (GNRs) of large widths, and also polymers with
large unit cells, the number of degrees of freedom involved in the
problem may impose severe limitations on the problems which can be
solved computationally. On the other hand the main advantage of the
effective $\pi$-electron model Hamiltonians is that they explicitly
deal only with the $\pi$ electrons, thereby reducing the degrees
of freedom considerably, and, thus allowing the simulation of much
larger systems as compared to the \emph{ab initio} approaches. Their
disadvantage, of course, is that they are semiempirical in nature,
and, therefore, the parameters involved in them are often determined
using the spectroscopic data of a suitable model system. Nevertheless,
when calculations on very large systems need to be performed, often
it is virtually impossible to use the \emph{ab initio} approaches,
and, therefore, model Hamiltonians provide an attractive alternative.
Even for smaller systems, model Hamiltonians allow us to understand
the underlying physics in simple terms, therefore, aforesaid $\pi$-electron
approaches are very popular when it comes to calculations of the electronic
structure of graphene and its nanostructures.

Among the effective $\pi$-electron approaches, the TB model (called
the H\"uckel model in the chemistry literature) is the simplest,
but it does not incorporate the effect of electron-electron (e-e)
repulsion. The Hubbard model and its extended versions go beyond the
TB approach, and incorporate the short-range parts of the Coulomb
repulsion such as the on-site, and the nearest-neighbor interactions,
respectively. In chemistry literature it is well-known that in $\pi$-electron
systems such as various aromatic molecules and conjugated polymers,
the role of long-range e-e interactions cannot be ignored when it
comes to their electronic structure\citet{salem-book,sumit-review,barford-book}.
And, indeed, Pariser-Parr-Pople (P-P-P) model Hamiltonian\citet{PPP},
which is an effective $\pi$-electron Hamiltonian incorporating long-range
e-e interaction, has been used with considerable success in describing
the physics of such systems\citet{salem-book,sumit-review,barford-book}.
Thus, it is logical to conclude that such e-e interactions will also
be important in understanding the physics of the graphene-based materials.
In recent years, our group and collaborators\citet{shuklamazumdar83,shuklaPRB-62,shukla-synthmet116,shukla-ppv-prb65,shuklaPRB-67,shukla-ppv-synth,shuklaCP-300,shuklaPRB-69,sony-pdpa,priya-pdpa-synth,priya-acene-synth,priya-acene-prb,priya-acene-jcp},
along with numerous other groups\citet{salem-book,sumit-review,barford-book,jug-ppp,ppp-soos,ppp-bredas,ppp-ramasesha,ppp-barford,ppp-sumit},
have used the P-P-P model to study the electronic structure and optical
properties of conjugated molecules and oligomers. For finite $\pi$-conjugated
systems, in our group, we have developed a general-purpose HF program
employing the P-P-P model, which is available to anyone for scientific
work\citet{sony-cpc}. However, GNRs and some $\pi$-conjugated polymers
are believed to be quasi-one-dimensional systems, which need to be
studied in the infinite-length limit. Thus, in order to study their
electronic structure and related properties, using the P-P-P model
(or any other Hamiltonian), one needs to impose periodic boundary
conditions, and perform infinite lattice sums to account for their
infinite extent, which, our earlier computer program\citet{sony-cpc}
lacks. With this aim in mind, we recently developed a P-P-P model-based
computer program, which solves both the restricted HF (RHF) and the
unrestricted HF (UHF) equations for one-dimensional periodic systems,
and used it to study the electronic structure and optical properties
of mono-layer and multilayer GNRs of various kinds\citet{gundra1,gundra2}.
The ability to solve the UHF equations allows us to explore magnetic
properties of polymers and GNRs. The aim of the present paper is to
describe the computer program in detail, and make it available for
use by anyone interested in the physics of GNRs and $\pi$-conjugated
polymers. The program is capable of computing the total energy, the
band structure, the density of states (DOS), and also the interband
optical absorption spectrum in form of the frequency-depdendent dielectric
response tensor, both with and without an external homogeneous electric
field. The fact that our program can solve the HF equations in the
presence of an external electric field allows the user to explore
gated configurations of polymers and GNRs. In this work, we demonstrate
the capabilities of our program by performing calculations on polymers
\emph{trans}-polyacetylene ($t$-PA), poly-\emph{para}-phenylene (PPP),
armchair GNRs (AGNRs), and zigzag GNRs (ZGNRs) of various widths.
We also note that Nakada \emph{et al}.\citet{ppp-nakada} reported
a P-P-P model based band structure calculation of both AGNRs and ZGNRs
several years back. 

The remainder of the paper is organized as follows. In section \ref{sec:theory}
we briefly review the theory associated with the P-P-P model Hamiltonian.
Next, in section \ref{sec:program} we discuss the general structure
of our computer program, and also describe its constituent subroutines.
In section \ref{sec:install} we briefly describe how to install the
program on a given computer system, and to prepare the input files.
Results of various example calculations using our program are presented
and discussed in section \ref{sec:results}. Finally, in section \ref{sec:Conclusions},
we present our conclusions, as well as discuss possible future directions.

\section{Theory}

\label{sec:theory}

In this section we briefly discuss the theory behind our computer
program.

\subsection{Pariser-Parr-Pople Hamiltonian}

\label{sub:pppham}

The P-P-P model Hamiltonian\citet{PPP}, with one $\pi$-electron
per carbon atom, is given by 
\begin{eqnarray}
H=\sum_{i,\sigma}\epsilon_{i}c_{i\sigma}^{\dagger}c_{i\sigma}-\sum_{i,j,\sigma}t_{ij}(c_{i\sigma}^{\dagger}c_{j\sigma}+c_{j\sigma}^{\dagger}c_{i\sigma})+\nonumber \\
U\sum_{i}n_{i\uparrow}n_{i\downarrow}+\sum_{i<j}V_{ij}(n_{i}-1)(n_{j}-1)\label{eq:ham-ppp}
\end{eqnarray}
 where $\epsilon_{i}$ represents the site energy associated with
the $i$th carbon atom, $c_{i\sigma}^{\dagger}$ creates an electron
of spin $\sigma$ on the $p_{z}$ orbital of atom $i$, $n_{i\sigma}=c_{i\sigma}^{\dagger}c_{i\sigma}$
is the number of electrons with the spin $\sigma$, and $n_{i}=\sum_{\sigma}n_{i\sigma}$
is the total number of electrons on atom $i$. The parameters $U$
and $V_{ij}$ are the on-site and long-range Coulomb interactions,
respectively, while $t_{ij}$ is the one-electron hopping matrix element.
On setting $V_{ij}=0$ (with $U\neq0)$, the Hamiltonian reduces to
the Hubbard model, while on setting both $U=0$ and $V_{ij}=0$, the
tight-binding (TB) model is obtained. The parametrization of Coulomb
interactions is Ohno like\citet{ohno},
\begin{equation}
V_{i,j}=\frac{U}{\kappa_{i,j}(1+0.6117R_{i,j}^{2})^{1/2}}\;\mbox{,}\label{eq-ohno}
\end{equation}
where, $\kappa_{i,j}$ depicts the dielectric constant of the system
which can simulate the effects of screening, and $R_{i,j}$ is the
distance in \AA{}~ between the $i$-th and the $j$-th carbon atoms.
\textcolor{black}{In our earlier work on GNRs\citet{gundra1}, we
used the }\textcolor{black}{\emph{ab-initio}}\textcolor{black}{{} GW
band structure of mono layer AGNR-12 (AGNR-$N_{A}$, denotes an AGNR
with $N_{A}$ dimer lines across the width) reported by Son }\textcolor{black}{\emph{et
al.}}\textcolor{black}{\citet{Son} to obtain a set of {}``modified
screened Coulomb parameters,'' with $U=6.0$ eV and $\kappa_{i,j}=2.0$
($i\neq j)$ and $\kappa_{i,i}=1$, which are slightly different from
the screened parameters reported initially by Chandross and Mazumdar\citet{chandross},
with $U=8.0$ eV and $\kappa_{i,j}=2.0$ ($i\neq j)$ and $\kappa_{i,i}=1$,
aimed at describing the optical properties of phenyl-based polymers
within the P-P-P model. The modified screened parameters provided
good agreement between our HF band gaps, and those obtained by the
GW method for a few AGNRs, however, for ZGNRs the agreement was not
good\citet{gundra1}. In this work, we examine the issue of the choice
of Coulomb parameters in a critical manner, and conclude that no single
set of parameters gives uniformly good agreements between our results
and the GW results for all types of GNRs. In section \ref{sec:results},
where this issue is investigated, we find that a set of parameters
which provides excellent agreement between ours and GW results for
a class of GNRs, may simply fail to reproduce such agreement for another
class of GNRs. In other words, the choice of Coulomb parameters which
will lead to good agreement between our HF results, and the }\textcolor{black}{\emph{ab
initio}}\textcolor{black}{{} GW ones, depends upon the geometries of
the GNRs in question. Thus, after trying a number of Coulomb parameters,
and in the absence of any experimental data on the band gaps of GNRs,
we have decided it is best to use the original screened parameters
of Chandross and Mazumdar\citet{chandross}, with $U=8.0$ eV and
$\kappa_{i,j}=2.0$ ($i\neq j)$ and $\kappa_{i,i}=1$ for all the
GNRs.} But, we would like to emphasize that the user has the freedom
to choose a different set of Coulomb parameters as per the requirements,
and, we encourage such experimentation.

\subsection{Unrestricted Hartree-Fock Equations}

\label{sub:uhfeq}

We have implemented the RHF and the UHF methods within the P-P-P model,
using the standard linear combination of atomic orbitals (LCAO) formalism.
We shall review the basics of the formalism for the UHF method (also
known as the Pople-Nesbet formalism\citet{szabo}), from which the
corresponding equations for the RHF method can be easily deduced.
In this approach, the $n-$th Bloch orbital of the system corresponding
to the spin up electrons ($\alpha/\beta$ will denote spin up/down
electrons) is expressed as a linear combination of $m$ basis functions
per unit cell, 

\begin{equation}
\psi_{n}^{(\alpha)}(k)=\sum_{\mu=1}^{m}C_{\mu n}^{(\alpha)}(k)\phi_{\mu}(k)\label{eq:blochalpha}
\end{equation}

where $C_{\mu n}^{(\alpha)}(k)$'s represent the linear expansion
coefficients, to be determined at a set of $k-$points in the first
Brillouin zone (BZ), and the $\mu$-th Bloch function $\phi_{\mu}(k)$
is given by

\begin{equation}
\phi_{\mu}(k)=\frac{1}{\sqrt{N}}\sum_{j}e^{ikR_{j}}\phi_{\mu}(r-R_{j})\label{eq:blochbasis}
\end{equation}

where $N\rightarrow\infty$ is the total number of unit cells in the
system, $\phi_{\mu}(r-R_{j})$ is the atomic orbital (AO) ($p_{z}$
orbital mentioned in section \ref{sub:pppham}) located in the $j^{th}$
unit cell defined by the lattice vector $R_{j}$. The definition of
the Bloch orbital corresponding to the spin down electrons will be
identical, with $\alpha$ replaced by $\beta$ in Eq. \ref{eq:blochalpha}.
Because, in the P-P-P model, the basis functions $\phi_{\mu}(r-R_{j})$
are assumed to form an orthonormal set, the UHF equation for up-spin
orbitals can be written in the matrix form as

\begin{equation}
F^{\alpha}(k)C_{n}^{\alpha}(k)=\epsilon_{n}^{\alpha}(k)C_{n}^{\alpha}(k)\label{eq:UHF-alpha}
\end{equation}

where, for a given $k$ value, $F^{\alpha}(k)$ represents the Fock
matrix for the up-spin electrons, $C_{n}^{\alpha}(k)$ represents
the corresponding $C_{\mu n}^{(\alpha)}(k)$ coefficents, arranged
in form of a column vector, and $\epsilon_{n}^{\alpha}(k$) denotes
the band eigenvalue. The Fock operator for electrons of up spin is
given by

\begin{equation}
F^{\alpha}(k)=h(k)+(J^{\alpha}(k)+J^{\beta}(k)-K^{\alpha}(k))\label{eq:fockalpha}
\end{equation}
where $h(k)$ denotes the Fourier transform of the one-electron parts
of the P-P-P Hamiltonian (\emph{cf}. Eq. \ref{eq:ham-ppp}), $J^{\alpha}(k)/K^{\alpha}(k)$
are the Coulomb/exchange integrals for the up spin electrons, obtained
by Fourier transforming their real-space counterparts
\begin{eqnarray}
h_{\mu\nu}(k) & = & \sum_{j=-\infty}^{\infty}e^{ikR_{j}}h_{\mu\nu}(R_{j}),\label{eq:hk}\\
J_{\mu\nu}^{\alpha}(k) & = & \sum_{j=-\infty}^{\infty}e^{ikR_{j}}J_{\mu\nu}^{\alpha}(R_{j}),\label{eq:jk}\\
K_{\mu\nu}^{\alpha}(k) & = & \sum_{j=-\infty}^{\infty}e^{ikR_{j}}K_{\mu\nu}^{\alpha}(R_{j}).\label{eq:kk}
\end{eqnarray}

Above $h_{\mu\nu}(R_{j})$ is the one-electron part of the P-P-P Hamiltonian,
and $J_{\mu\nu}^{\alpha}(R_{j})/K_{\mu\nu}^{\alpha}(R_{j})$ denote
the Coulomb/exchange integrals in the real space, defined as

\begin{equation}
J_{\mu\nu}^{\alpha}(R_{j})=\sum_{\sigma=1}^{m}\sum_{\lambda=1}^{m}\sum_{k=-\infty}^{\infty}D_{\sigma\lambda}^{\alpha}(R_{k})\sum_{l=-\infty}^{\infty}\langle\mu(o)\sigma(R_{l})|\frac{1}{r_{12}}|\nu(R_{j})\lambda(R_{l}+R_{k})\rangle,\label{eq:jalpha}
\end{equation}
and 
\begin{equation}
K_{\mu\nu}^{\alpha}(R_{j})=\sum_{\sigma=1}^{m}\sum_{\lambda=1}^{m}\sum_{k=-\infty}^{\infty}D_{\sigma\lambda}^{\alpha}(R_{k})\sum_{l=-\infty}^{\infty}\langle\mu(o)\sigma(R_{l})|\frac{1}{r_{12}}|\lambda(R_{l}+R_{k})\nu(R_{j})\rangle,\label{eq:kalpha}
\end{equation}
where the expression for a general two-electron repulsion integrals
is 

\begin{eqnarray}
\langle\mu(R_{i})\sigma(R_{j})|\frac{1}{r_{12}}|\nu(R_{k})\lambda(R_{l})\rangle & = & \int\int\phi_{\mu}({\bf r}_{1}-R_{i})\phi_{\nu}({\bf r}_{1}-R_{k})r_{12}^{-1}\nonumber \\
 & \times & \phi_{\sigma}({\bf r}_{2}-R_{j})\phi_{\lambda}({\bf r}_{2}-R_{l})d^{3}{\bf r}_{1}d^{3}{\bf r}_{2},\label{eq:twoint}
\end{eqnarray}

and the density matrix for the up-spin electrons, $D_{\mu\nu}^{\alpha}(R_{j})$,
is given by 
\begin{equation}
D_{\mu\nu}^{\alpha}(R_{j})=\frac{1}{\Delta}\int\sum_{n=1}^{n_{\alpha}}C_{\mu n}^{\alpha*}(k)C_{\nu n}^{\alpha}(k)e^{ikR_{j}}dk,\label{eq:denmat}
\end{equation}

where the integral over $k$ is performed over the one-dimensional
(1D) BZ of length $\Delta$, and $n_{\alpha}$ denotes the number
of up-spin electrons per unit cell. Above we have given the explicit
expressions of various quantities for their up-spin components only,
because the expressions for the down-spin components can be obtained
simply by interchanging $\alpha$ and $\beta$. The total energy per
unit cell of a given system is computed using the real-space expression
\begin{eqnarray}
E_{cell} & = & \sum_{j}\sum_{\mu,\nu}\{D_{\mu\nu}(R_{j})h_{\mu\nu}(R_{j})+\frac{1}{2}D_{\mu\nu}^{\alpha}(R_{j})(J_{\mu\nu}^{\alpha}(R_{j})-K_{\mu\nu}^{\alpha}(R_{j}))\nonumber \\
 &  & +\frac{1}{2}D_{\mu\nu}^{\beta}(R_{j})(J_{\mu\nu}^{\beta}(R_{j})-K_{\mu\nu}^{\beta}(R_{j}))+D_{\mu\nu}^{\beta}(R_{j})J_{\mu\nu}^{\alpha}(R_{j})\},\label{eq:total-energy}
\end{eqnarray}

where $D_{\mu\nu}(R_{j})=D_{\mu\nu}^{\alpha}(R_{j})+D_{\mu\nu}^{\beta}(R_{j})$,
denotes the total density matrix of the system. The expressions of
the two-electron integrals appearing in Eqs. \ref{eq:jalpha} and
\ref{eq:kalpha} are of the most general type, however, in case of
the P-P-P model (\emph{cf}. Eq. \ref{eq:ham-ppp}) only density-density
type of e-e repulsion terms are included, which implies

\begin{equation}
\langle\mu(R_{i})\sigma(R_{j})|\frac{1}{r_{12}}|\nu(R_{k})\lambda(R_{l})\rangle=\delta_{\mu\nu}\delta_{\sigma\lambda}\delta_{R_{i}R_{k}}\delta_{R_{j}R_{l}}V_{\mu(o)\lambda(R_{j}-R_{i})}\label{eq:twoint-ppp}
\end{equation}
where $V_{\mu(0)\lambda(R_{j}-R_{i})}$ implies that expression is
calculated using Eq. \ref{eq-ohno}, assuming that the $\mu$-th basis
function is located in the reference unit cell while the $\lambda$-th
basis function is in the cell with location $R_{j}-R_{i}$. After
the simplification of Eq. \ref{eq:twoint-ppp}, evaluation of $J_{\mu\nu}^{\alpha}(R_{j})$
and $K_{\mu\nu}^{\alpha}(R_{j})$ becomes quite easy: (a) in Eq. \ref{eq:jalpha}
only an infinite lattice sum over $R_{l}$ needs to be performed,
which is done by including a large number of terms, and (b) in Eq.
\ref{eq:kalpha} both the sums for $R_{k}$ and $R_{l}$ reduce to
one term each. The convergence of our calculations with respect to
these lattice sums was tested extensively.

The UHF equations of the system, leading to the band structure ($\epsilon_{n}^{(\alpha)}(k)$/$\epsilon_{n}^{(\beta)}(k)$),
and the correpsponding Bloch orbitals, are obtained by solving Eq.\ref{eq:UHF-alpha},
and its $\beta$-spin counterpart, by iterative diagonalization technique
at a set of $k$-points, until the total energy per cell of the system
(\emph{cf}. Eq. \ref{eq:total-energy}) converges. During the self-consistent
HF iterations, the integration over the BZ is performed using the
Gauss-Legendre quadrature technique as suggested by André \emph{et
al}.\citet{hf-pol}, with the additional flexibility that the number
of points used for the quadrature can be chosen by the user.

In order to perform calculations in the presence of a static external
electric field to simulate the gate bias, one can solve the HF equations
using a modified Fock operator under the electric dipole approximation
by introducing the corresponding term containing the uniform electric
field $\mathbf{E}$. The modified Fock operator of the system is then
given by

\begin{equation}
F_{efield}^{\alpha}=F^{\alpha}-\mathbf{\boldsymbol{\mu}.E}=F^{\alpha}+|e|\mathbf{E.r}\,,\label{eq:PPP-efield-Ham}
\end{equation}
where $F^{\alpha}$ is the unperturbed Fock operator for the up-spin
electrons in the absence of the electric field, $e$ represents the
electronic charge, $\boldsymbol{\mu}=-e\mathbf{r}$, is the dipole
operator, and $\mathbf{r}$ is the position operator for which the
usual diagonal representation is employed.

\subsection{Density of states}

The density of states (DOS) is obtained using the well-known expression 

\begin{equation}
\rho(\epsilon)=C\sum_{i}\int e^{-(\epsilon-\epsilon_{i}(k))^{2}/2\gamma^{2}}dk\label{eq:dos}
\end{equation}
where $\epsilon$ is energy at which DOS is computed, $\epsilon_{i}(k)$
is the energy of $i$-th orbital at a given $k$ point,$\gamma$ is
the broadening parameter, and $C$ includes the rest of the constants.
The integration over $k$ is performed over the 1D BZ, and summation
over $i$ includes all the Bloch orbitals of the system. In our calculations
we set $C=1$ to obtain the DOS in the arbitrary units.

\subsection{Theory of optical absorption}

The optical absorption spectrum of the incident radiation polarized
in $x$ or $y$ direction is computed in the form of the corresponding
components of the imaginary part of the dielectric constant tensor,
i.e., $\epsilon_{ii}(\omega)$, using the standard formula
\begin{equation}
\epsilon_{ii}(\omega)=C\sum_{v,c}\int\frac{|\langle c(k)|p_{i}|v(k)\rangle|^{2}}{\{(E_{cv}(k)-\hbar\omega)^{2}+\gamma^{2}\}E_{cv}^{2}(k)}dk,\label{eq:eps2}
\end{equation}
where $p_{i}$ denotes the momentum operator in the $i$-th Cartesian
direction, $\omega$ represents the angular frequency of the incident
radiation, $E_{cv}(k)=\epsilon_{c}(k)-\epsilon_{v}(k)$, with $\epsilon_{c}(k)\:(\epsilon_{v}(k))$
being the conduction band (valence band) eigenvalues of the Fock matrix,
$\gamma$ is the line width, while $C$ includes rest of the constants.
Assuming that the valence band eigen state $|v(k)\rangle$ is expressed
as (\emph{cf}. Eq. \ref{eq:blochalpha}, ignoring the spin orientation)

\[
|v(k)\rangle=\sum_{\mu}C_{\mu v}(k)|\chi_{\mu}(k)\rangle,
\]
 with a similar expression for the conduction band eigen states $|c(k)\rangle$.
The momentum matrix elements $\langle c(k)|p_{i}|v(k)\rangle$ needed
to compute $\epsilon_{ii}(\omega)$, for a 1D periodic system, can
be calculated using the formula,\citet{opt-mat-el-2} 

\begin{eqnarray}
\langle c(k)|p_{i}|v(k)\rangle & =\delta_{i,1} & \frac{m_{0}}{\hbar}\sum_{\mu,\nu}C_{\nu c}^{*}(k)C_{\mu v}(k)\frac{\partial}{\partial k}H_{\nu\mu}(k)\nonumber \\
 &  & +\frac{im_{0}(\epsilon_{c}(k)-\epsilon_{v}(k))}{\hbar}\sum_{\mu,\nu}C_{\nu c}^{*}(k)C_{\mu v}(k)d_{\mu\nu}^{(i)},\label{eq:p-matel}
\end{eqnarray}
where $\delta_{i,1}$ implies that the term is nonzero when $i$ denotes
the periodicity direction ($x$ direction), $m_{0}$ is the free-electron
mass, $\frac{\partial}{\partial k}H_{\nu\mu}(k)$ represents the derivative
of the Hamiltonian (Fock matrix, in the present case) with respect
to $k$, $d_{\mu\nu}^{(i)}$ denotes the matrix elements of the $i$-th
component of the position operator ${\bf \textbf{\ensuremath{\mathbf{d}}}}$
defined with respect to the reference unit cell, and accounts for
the so-called intra-atomic contribution\citet{opt-mat-el-2}. Note
that Eq. \ref{eq:p-matel} can also be used to compute the matrix
element $\langle c(k)|p_{y}|v(k)\rangle$ needed to calculate the
absorption spectrum for the $y$-polarized light for GNRs (which are
periodic only in the $x$ direction), by setting the first term on
its right hand side to zero, and retaining the contribution only of
the second term. In these calculations, $\frac{\partial}{\partial k}H_{\nu\mu}(k)$
was computed numerically, while the usual diagonal representation
was employed for the ${\bf d}$ operator. Furthermore, we set $C=1$
in all the cases to obtain the absorption spectra in arbitrary units.

\section{Description of the Program}

\label{sec:program}

Theoretical formalism described in the preceding section has been
implemented numerically in a computer program called {}``ppp\_bulk.x''
using the Fortran 90 (F90) programming language. Advanced features
of F90, such as the dynamic memory allocation, modules, etc,. have
been used to improve the efficiency of the code. Architecture independent
numerical precision is used for portability of the code across different
computer architectures. This program is capable of doing both the
tight binding as well as P-P-P model calculations. Using a small number
of input parameters such as the positions of the atoms in the unit
cell, lattice translation vector, hopping and P-P-P Coulomb parameters
etc., ppp\_bulk.x determines the band structure, density of states,
joint density of states, optical and electro absorption spectrum for
1D periodic $\pi$-conjugated systems. As mentioned earlier, the lattice
sums are performed in real space by including a large number of unit
cells, and integration along the BZ was performed using the Gauss-Legendre
quadrature approach\citet{hf-pol}. The convergence of SCF iterations
is slow for systems with large number of electrons, therefore, we
have also implemented the method of damping to speed up the convergence.

Our computer code consists of the main program, and various subroutines.
Optionally, the user can link to the LAPACK/BLAS libraries, whose
diagonalization routines can be used by our program. In the following
we briefly describe the main program, as well as each subroutine/function.

\subsection{Module MTYPES}

In this module the precision of REAL variables used throughout the
program is defined. This will facilitate machine independent precision
for REAL variables.

\subsection{Module MCOMMONDATA}

Global data shared by several routines in the program is defined in
this module.

\subsection{Main program PPP\_BULK}

This is the main program of our package for performing electronic
structure calculations within a semiempirical TB formalism for 1D
periodic systems. It calls other subroutines to accomplish various
tasks.

\subsection{Subroutine INPUT}

This routine reads input data such as which Hamiltonian to use, its
parametrization, total number of atoms in the unit cell, their Cartesian
coordinates, 1D lattice constant, and\emph{ }number of $k$-points
($n_{k}$) for BZ sampling\emph{ etc}. Besides, this subprogram reads
the options to perform various types of calculations such as tight
binding, RHF, UHF \emph{etc}. This subroutine also reads the components
of external electric field in the units of V/\AA, if calculations
for a gated configuration need to be performed. Alternatively, by
calling other routines, one can generate coordinates of some important
structural units such as AGNRs and ZGNRs of different widths to facilitate
an easy realization of the GNR under consideration. In addition, various
arrays are allocated dynamically and deallocated after the desired
task is completed.

\subsection{Subroutine GNR\_RATOM}

This routine generates the coordinates of various atoms in the unit
cell of the given GNR, based upon the user specified data consisting
of the type of GNR (AGNR or ZGNR), its width, and the nearest-neighbor
bond length. It also computes the lattice translation vector.

\subsection{Subroutine ERROR}

This routine writes out a fatal error message to a user specified
logical unit, and stops the execution of the code.

\subsection{Subroutine SORT}

\textcolor{black}{This routine, adapted from Ref.\citet{nrf77},}
generates an array in which distances between different pairs of atoms
in the system are stored in the ascending order.

\subsection{Subroutine GET\_NN}

This routine uses the array generated by subroutine SORT to computes
the distance between pair of atoms which are nearest neighbors (NN),
second NN, third NN and so on depending the number of unique hoppings
defined.

\subsection{Subroutine get\_NTij}

This subroutine computes the total number of hopping matrix elements
connecting various atomic sites in the system.

\subsection{Subroutine HOPPING}

This routine generates the hopping matrix elements connecting various
sites. Hopping matrix elements are assigned for a pair of atoms depending
on the distance between them. The unique hopping matrix elements are
defined by user in the input file starting from NN atoms, followed
by the second NN atoms, and so on.

\subsection{Subroutine IJPK}

This subroutine is used to pack the row index $i$ and column index
$j$ of an element of the upper triangle of a real symmetric matrix
into an integer corresponding to its location in a 1D array.

\subsection{Subroutine INVIJPK}

The task of this subroutine is just the reverse of the subroutine
IJPK, \emph{i.e.}, it is used to unpack the integers $i$ (row index)
and $j$ (column index) from the location of the corresponding matrix
element of an upper-triangular symmetric matrix packed in a 1D array.

\subsection{Subroutine CHEKEDGE}

This subroutine is used to find whether a given site lies on the edge
or in the interior of the system. It calculates the $NN$ for the
given site, and if $NN=$2 then it is regarded as an edge site, and
if $NN=3$ then it is a site in the interior of the system.

\subsection{Subroutine FILOPN}

This subroutine is meant for opening a file which may either be new
or old.

\subsection{Subroutine FILCLS}

This subroutine is meant for closing an already open file.

\subsection{Function DOTPD}

This function calculates the dot product between two given vectors.

\subsection{Subroutine GETNA}

Assuming that a given lattice vector (${\bf r}$) is in the form ${\bf r}=n{\bf a}$
and finds the integer $n$, where ${\bf a}$ is the primitive vector
of the lattice.

\subsection{Subroutine PPP\_PARA}

This subroutine generates parameters associated with the P-P-P Hamiltonian,
as per the user choice.

\subsection{Subroutine PRINTR}

This subroutine prints the coordinates of the unit cell in an output
file called 'unitcell.xsf', which can viewed using the visualization
packages such as xcrysden\citet{xcrysden}. Additionally it also generates
an output file called 'system.xsf' in which coordinates of atoms in
several unit cells are printed for viewing the periodic system under
consideration.

\subsection{Subroutine CELL\_DRV}

This is the driver routine for generating the unit cell related data
by calling another subroutine named 'cell\_1d'.

\subsection{Subroutine CELL\_1D}

Starting with the primitive cell and the lattice vector related data,
it generates the coordinates of all the cells and atoms included in
the calculations. This data is useful when lattice sums, to account
for the long-range Coulomb interactions, are performed.

\subsection{Subroutine MATEL\_R}

This is the master routine meant for generating the one- and two-electron
matrix elements in the real space, with, or without, the external
electric field. This is done based upon the data specified by the
user in the input file such as the Hamiltonian under consideration,
Coulomb parameters to be used (if any), hopping matrix elements connecting
various sites, \textit{etc}.

\subsection{Subroutine READ\_Ri}

This routine reads the coordinates of atoms in different unit cells,
generated by the routine cell\_1d, for performing real space lattice
sums.

\subsection{Subroutine Makntpq}

This routine arranges the unit cells in an order required for performing
real-space lattice sums. The reference unit cell is numbered '$0$',
the unit cells to the right of reference unit cells are identified
labeled with positive integers, and the unit cells to its left are
labeled with negative integers.

\subsection{Subroutine Maktmat}

This routine stores the hopping matrix elements in a translationally
invariant format . In this format each hopping element is stored as
$t_{ij}(ipq)$, where orbital $i$ is assumed to be in the cell at
location $ipq$, while orbital $j$ is in the reference cell.

\subsection{Subroutine NUCNUC}

This subroutine computes real space nucleus-nucleus repulsion term
for the P-P-P model.

\subsection{Subroutine ELE\_NUC}

This subroutine computes real space electron-nucleus repulsion term
for the P-P-P model.

\subsection{Subroutine COULOMB}

This subroutine computes the long-range Coulomb part of the e-e repulsion
term $J_{\mu\nu}(R_{j}$) (\textit{cf}. Eq. \ref{eq:jalpha}) for
the P-P-P model.

\subsection{Subroutine EXCHANGE}

This subroutine computes the long-range exchange part of the e-e repulsion
term $K_{\mu\nu}(R_{j})$ (\textit{cf}. Eq. \ref{eq:kalpha}) corresponding
to the P-P-P Hamiltonian.

\subsection{Subroutine gen\_Kmesh}

This subroutine creates $k$-points in the positive part of the first
BZ of a 1D system, between the limits $0$ and $\pi$. Various $k$-points
are non-equidistant, and chosen in accordance with the Gauss-Legendre
quadrature. The BZ integration is performed for the dimensionless
variable $ka$, $a$ being the lattice constant.

\subsection{Subroutine GAULEG}

This subroutine generates the roots and weights needed for the Gauss-Legendre
quadrature meant for BZ integration.

\subsection{Subroutine SCF\_DRV}

This is the driver routine for carrying out the iterative self-consistent
field (SCF) calculations.

\subsection{Module MSCF\_VAR}

This module defines the variables common to routines that perform
SCF iterations. In addition, subroutines for allocation and deallocation
of several arrays are also defined in this module.

\subsection{Subroutine SCF\_RHF }

This subroutine solves the RHF equations for the system under consideration
in a self-consistent manner, using the iterative diagonalization procedure
at each $k$-point, and returns the canonical SCF orbitals, their
eigenvalues, and the total energy per unit cell of the system. The
arrays which are needed during the calculations are allocated before
the calculations begins, and are deallocated upon its completion.
Before the first iteration, H\"uckel model Hamiltonian is diagonalized
to obtain a set of starting orbitals. Subsequently, the Fock matrix
corresponding to those orbitals is constructed, and diagonalized.
The process is repeated until the self-consistency is achieved. Depending
upon the choice of the user, the eigenvalues and eigen vectors can
be obtained using either the ZHPEV subroutine from the LAPACK/BLAS
library, or the inbuilt HOUSEH\_C subroutine\citet{househ}, based
upon the Householder diagonalization scheme.

\subsection{Subroutine SCF\_UHF }

This subroutine is exactly the same in its logic and structure as
the previously described SCF\_RHF, except that the task of this routine
is to solve the UHF equation for the system under consideration. Different
Fock matrices for the $\alpha$ and the $\beta$ spin are constructed
and diagonalized in each iteration, until the self-consistency is
achieved. Eigenvalues and eigen vectors are computed using the routine
ZHPEV/HOUSEH\_C. The iterations are stopped once the total UHF energy
of the system converges to within a user defined threshold.

\subsection{Subroutine COMMUL}

This subroutine computes the product of two complex numbers, and returns
the real and imaginary parts of the product, separately.

\subsection{Subroutine GRADFK}

This subroutine finds the $k$-space derivative of Fock matrix ($\frac{\partial F(k)}{\partial k}$),
using the central difference formula. Recall that $\frac{\partial F(k)}{\partial k}$
is needed to evaluate the momentum matrix elements (\emph{cf}. Eq.
\ref{eq:p-matel}) required for computing the linear optical absorption
spectrum.

\subsection{Subroutine BAND}

This subroutine computes the band structure of the system by diagonalizing
the converged Fock matrix at a large number of uniformly-spaced set
of $k$ points. For the purpose, the Fock matrix at those $k$ points
is obtained by Fourier transforming converged Fock matrix in the real
space. The Fock matrix in the $k$-space is diagonalized using the
subroutine ZHPEV/HOUSEH\_C to obtain its eigenvalues and eigen vectors.

\subsection{Subroutine PRINT\_Ek}

In this routine, the eigenvalues of the Fock matrix obtained in the
subroutine BAND are written in various ASCII files. If the calculation
was an RHF one, then the data is written in the file named 'bands.dat',
while for the UHF calculations the up-spin eigenvalues are written
in the file 'bandsUP.dat', and the down-spin ones in the file 'bandsDOWN.dat'
. These files can be used for plotting the band structure using the
standard graphics packages such as xmgrace\citet{xmgrace} or gnuplot\citet{gnuplot}.

\subsection{Subroutine DOS }

This subroutine computes the density of states (DOS) and writes it
in an ASCII file named 'dos.dat'. The input to the routine consists
of the energy windows, and the broadening parameter (\emph{cf}. Eq.
\ref{eq:dos}).

\subsection{Subroutine OPTICS}

This is the master routine meant for evaluating the optical absorption
spectrum. Several other subroutines are called in this routine to
perform specific tasks. The range of frequencies over which the spectrum
is to be computed, along with the line width, are read from the input
file. The calculated spectrum is written in output files called 'sigma\_x.dat'
for incident radiation polarized along $x$-axis, and 'sigma\_y.dat',
for incident radiation polarized along $y$-axis, respectively. Furthermore,
the joint-density of stats is written in the file named 'jdos.dat'.
The data in all these files can be visualized using packages xmgrace\citet{xmgrace}
or gnuplot\citet{gnuplot}. In addition, several output files named
'sigma\_mn.dat', where band-specific optical absorption spectra, due
to transition from the $m^{th}$ valence band to the $n^{th}$conduction
band, are also generated.

\subsection{Subroutine SINCOS}

This routine computes $\cos(kR_{j})$ and $\sin(kR_{j})$ for various
values of $k$ and $R_{j}$, needed for performing the Fourier transforms
of various quantities.

\subsection{Subroutine makGmat}

This routine constructs the electron repulsion part of Fock matrix
for the RHF calculations performed within the P-P-P model Hamiltonian.

\subsection{Subroutine makGmat\_UHF}

This subroutine is analogous to the routine makGmat, the only difference
being that it constructs the repulsion matrix separately for the up-
and the down-spin electrons, and is called when UHF (as against RHF)
calculations are needed.

\subsection{Subroutine FOUTRA}

This subroutine performs a Fourier transforms on a given complex operator,
either from the ${\bf k}$-space to the ${\bf r}$-space, or vice
versa, depending upon the value of the input variable 'iflag'. Because
the systems under consideration are 1D, the BZ integration is performed
only in the positive part of the BZ, as the contribution of the negative
part of the BZ is same as that from the positive part.

\subsection{Subroutine HOUSEH\_C}

This subroutine, which originally belongs to the EISPACK library\citet{househ},
is used for diagonalizing the Fock matrix at different $k$ points
to obtain its eigenvalues and eigenvectors. For the purpose, the Householder
diagonalization approach is utilized\citet{househ}.

\subsection{Subroutine DENK\_RHF}

This subroutine constructs the density matrix in the $k$ space, using
the Bloch orbitals for the closed-shell systems, assuming that the
orbitals are doubly occupied.

\subsection{Subroutine DENK\_UHF}

This subroutine performs the same function for UHF calculations, which
the routine DENK\_RHF performs for the RHF calculations. It generates
different density matrices for the orbitals with $\alpha$ and $\beta$
spins, and also obtains the total density matrix by adding them.

\subsection{Subroutine SYMMAT}

This subroutine multiplies the off-diagonal Fock matrix elements of
a real-symmetric matrix by a factor of $2$ to enforce the upper-triangular
nature of the matrix.

\subsection{Subroutine xMMEcv}

Using $\frac{\partial F(k)}{\partial k}$ computed in the subroutine
'GRADFK', this subroutine computes the momentum matrix element $\langle c(k)|p_{x}|v(k)\rangle$
defined in Eq.\ref{eq:p-matel} for incident radiation polarized along
the $x$-direction.

\subsection{Subroutine yMMEcv}

This subroutine computes the momentum matrix element $\langle c(k)|p_{y}|v(k)\rangle$
defined in Eq.\ref{eq:p-matel} for the incident radiation polarized
along the $y$-direction.

\subsection{Function LINESHAPE}

This function computes the Lorentzian line shape, inputs are line
width, frequency of the incident radiation, and the energy difference
between the conduction- and valence-band eigenvalues, at a given $k$-point.

\subsection{Subroutine SIGMA\_ABSORB}

This routine evaluates the optical absorption spectrum defined in
Eq.\ref{eq:eps2}.

\subsection{Subroutine printTimeDate}

This subroutines uses the Fortran 90 intrinsic routine DATE\_AND\_TIME
to print date and time.

\subsection{Subroutine bloch\_out}

This subroutine generates an output file named 'bloch.dat', in which
the eigenvectors are written in the binary format.

\section{Installation, input files, output files}

\label{sec:install}

We believe that the installation and execution of the program, as
well as preparation of suitable input files is fairly straightforward.
Therefore, we will not discuss these topics in detail here. Instead,
we refer the reader to the README file for the details related to
the installation and execution of the program. Additionally, the file
'\texttt{manual.pdf}' explains how to prepare a sample input file.
Several sample input and output files corresponding to various example
runs are also provided with this package.

\section{Results and Discussions}

\label{sec:results}

In this section we demonstrate the abilities of our code by presenting
results for some 1D $\pi$-conjugated systems such as $t$-PA, PPP,
\textcolor{black}{AGNR-11}, AGNR-14, \textcolor{black}{ZGNR-8 }(ZGNR-$N_{Z}$,
denotes a ZGNR with $N_{Z}$ zigzag lines across the width), and ZGNR-10.
\textcolor{black}{Unless otherwise specified, all the calculations
are performed using the screened parameters with $U=8.0$ eV and $\kappa_{i,j}=2.0$
($i\neq j)$ and $\kappa_{i,i}=1$. However, later on we examine the
influence of Coulomb parameters on the band structure of GNRs.} In
order to benchmark our code, first we use it to compute the total
energy per unit cell ($E_{cell}$) for $t$-PA and the polymer PPP,
and compare it with the results obtained using our earlier P-P-P code
meant for finite systems\citet{sony-cpc}, with the increasing system
size. We also analyze the convergence of the total energy/cell with
respect to the total number of $k$ points ($n_{k}$) used for the
BZ integration, number of cells considered for summation of exchange
integral ($n_{exc}$\emph{, cf.} Eq. \ref{eq:kalpha}), by means of
calculations on AGNR-6. Furthermore, we present our results for the
band structure, density of states, linear optical absorption spectrum
of some of the systems mentioned above. The calculations and analysis
of the electric filed driven half-metallicity of ZGNR-14 and the electro-absorption(EA)
spectrum of ZGNR-8 are also presented.

\subsection{Total energy}

The variation of total energy per unit cell, $E_{cell}$ for finite
fragments of t-PA and PPP with the increasing number of cells ($N$)
obtained using our earlier P-P-P code for the finite systems\citet{sony-cpc}
is tabulated in Table \ref{Table1}, while the $E_{cell}$ for the
same systems in the infinite polymer limit, obtained by the present
code ppp\_bulk.x, is presented in the last column of the table.\textcolor{green}{{}
}\textcolor{black}{The t-PA was considered in its dimerized configuration,
with the single (double) bond length of $1.45$\AA ($1.35$ \AA),
and the corresponding hopping matrix element to be $2.232$ eV ($2.568$
eV). In case of the polymer PPP, the intra-phenyl C-C bond length
was taken to be $1.4$ \AA\  with the corresponding hopping of $2.46$
eV, while the length of the single bond connecting neighboring phenyl
rings was assumed to be $1.54$ \AA, along with the associated hopping
of 2.23 eV. From table \ref{Table1} it is evident that, the there
is excellent agreement between the two sets of calculations, which
gives us confidence about the essential correctness of our code.}

\begin{table}
\caption{Variation of total energy (in eV) per unit cell ($E_{cell}$) of t-PA
and PPP with the number of unit cells ($N$) obtained using our earlier
P-P-P code for finite systems\citet{sony-cpc}, compared with the
result for the infinite polymer obtained by the present code (last
column). }

\begin{tabular}{|c|c|c|c|c|c|}
\hline 
System & $N=5$ & $N=10$ & $N=50$ & $N=100$ & Infinite Polymer (This work)\tabularnewline
\hline 
\hline 
t-PA & -3.20 & -3.30 & -3.38 & -3.39 & -3.40\tabularnewline
\hline 
PPP & -11.66 & -11.73 & -11.79 & -11.79 & -11.81\tabularnewline
\hline 
\end{tabular}\label{Table1}
\end{table}

\begin{figure}
\includegraphics[width=10cm]{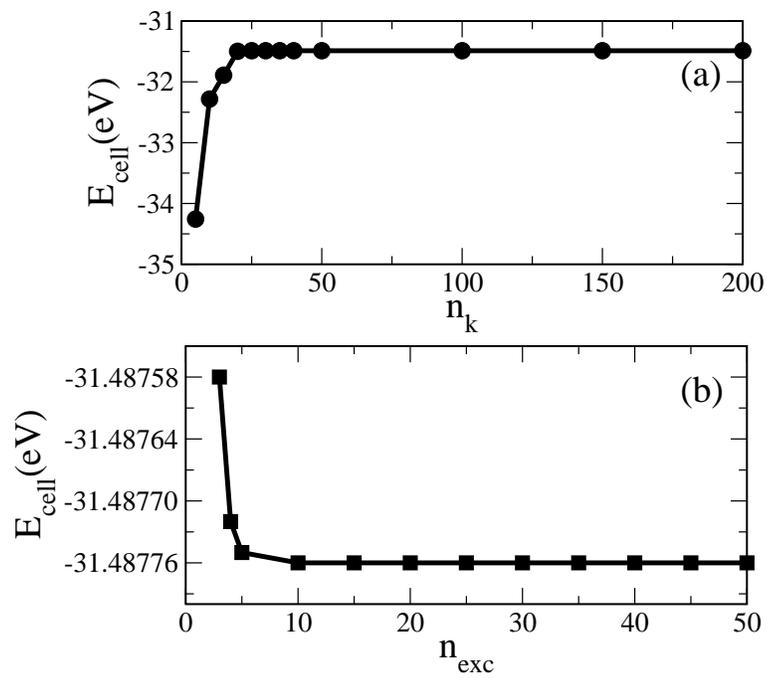}

\caption{Variation of total energy per unit cell of AGNR-6 with respect to:
(a) $n_{k}$, and (b) $n_{exc}$.}

\label{Flo:e_total}
\end{figure}

Next, we examine the convergence of $E_{cell}$ for AGNR-6 with the
increasing values of $k$ points ($n_{k}$) used for BZ integration
(Fig. \ref{Flo:e_total}(a)) and with the increasing values of the
number of unit cells in the exchange sum, $n_{exc}$ (Fig. \ref{Flo:e_total}(b)).
The nearest-neighbor (NN) hopping was chosen to be $t=2.7$ eV. From
Fig. \ref{Flo:e_total} it is obvious that: (a) $E_{cell}$ converges
at about $n_{k}=25$, and remains insensitive to further increase
in the value $n_{k}$, and (b) $E_{cell}$ converges rapidly with
$n_{exe}$ and convergence is achieved for $n_{exe}=10$. In both
cases we have considered about 10000 cells to evaluate the Coulomb
part of two electron integrals (\emph{c.f} Eq.\ref{eq:jalpha}). For
rest of the calculations we have chosen the values : $n_{k}=50$ and
$n_{exe}=24$. 

\subsection{Band structure and the Density of States}

Our code can also be used to perform band structure calculations,
along with the associated DOS using the TB model, as well as the P-P-P
model, employing both the RHF and the UHF methods. In what follows
below, we present the band structure and DOS for a few of the systems,
computed using our code.

\subsubsection{\emph{Trans}-polyacetylene}

\begin{figure}
\includegraphics[width=10cm]{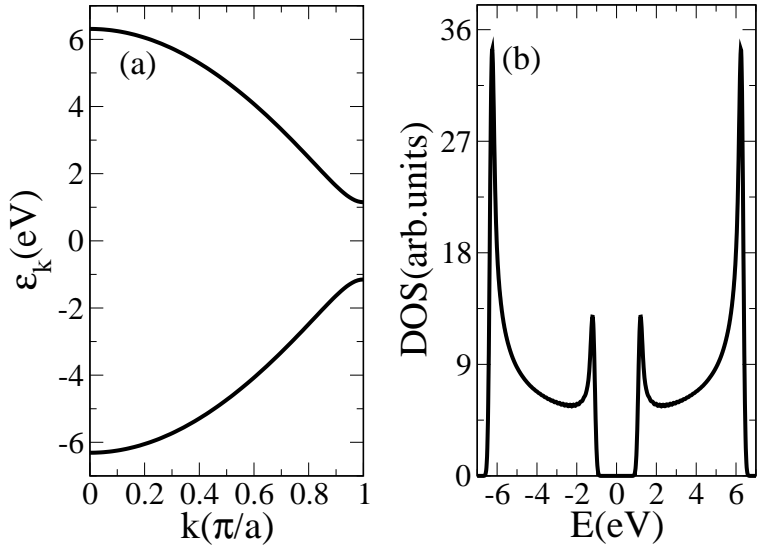}

\caption{(a) Band structure, and (b) DOS of dimerized t-PA obtained using the
P-P-P model, and the RHF method.}

\label{Flo:band-tpa}
\end{figure}

The band structure of the dimerized \emph{t}-PA obtained by the P-P-P-RHF
method is presented in Fig. \ref{Flo:band-tpa} (a), with the Fermi
energy ($E_{F}$) set to 0. The band gap occurs at $k=\pi/a$, and
is obtained to be 2.30 eV, a value in excellent agreement with the
the HOMO-LUMO gap of 2.31 eV obtained for a finite fragment of t-PA
consisting of 100 unit cells, computed using our earlier molecular
P-P-P code\citet{sony-cpc}. The vanishing DOS (Fig. \ref{Flo:band-tpa}(b))
around the Fermi energy characterizes the band gap, while the DOS
peaks near $E=\pm1.15$ eV and $E=\pm6.3$ eV are due to the van Hove
singularities at $k=\pi/a$ and $k=0$, respectively.

\subsubsection{Graphene Nanoribbons}

In this section we present the results of band structure calculations
on both AGNRs and ZGNRs. Furthermore, we also examine the electricity-driven
half-metallicity predicted in ZGNRs, by computing its band structure
in the presence of a transverse static electric field. However, first
we examine the influence of Coulomb parameters on the electronic structure
of GNRs. In all the calculations on the GNRs reported in the remainder
of this work, carbon-carbon nearest-neighbor distance was taken to
be $1.42$ \AA, and all the bond angles were assumed to be 120\textsuperscript{o}.
As far as the hopping matrix elements are concerned, for both AGNRs
and the ZGNRs, the nearest-neighbor hopping was taken to be 2.7 eV,
while, only for ZGNRs, a second nearest-neighbor hopping of $t'=0.27$
eV was also included.

\paragraph{Influence of Coulomb parameters}

\textcolor{black}{In our earlier work on GNRs\citet{gundra1}, we
proposed a set of {}``modified screened Coulomb parameters,'' with
$U=6.0$ eV and $\kappa_{i,j}=2.0$ ($i\neq j)$ and $\kappa_{i,i}=1$,
however, we did not present a detailed analysis of the parameter dependence
of our results. In Tables \ref{tab:AGNR-gaps} and \ref{tab-ZGNR-gaps}
we present the results of our P-P-P-HF calculations of the band gaps
of AGNRs and ZGNRs of varying widths as a function of the Coulomb
parameter $U$, while the screening defined by $\kappa_{i,j}=2.0$
($i\neq j)$ and $\kappa_{i,i}=1$ is employed. Furthermore, we also
compare our results to the }\textcolor{black}{\emph{ab initio}}\textcolor{black}{{}
DFT and GW results}\foreignlanguage{american}{\textcolor{black}{\citet{yang2}}}\textcolor{black}{. }

\begin{table}
\caption{Band gaps of different families of AGNR-$N_{A}$ ($N_{A}=3p,\;3p+1,\;3p+2,$\textcolor{red}{{}
}\textcolor{black}{($p$ $\ge1$, is an integer}) obtained using our
P-P-P RHF approach for $U=6.0$ eV and $U=8.0$ eV, compared with
the \emph{ab initio }DFT and GW results of Yang \emph{et al}.\citet{yang2}.}

\selectlanguage{american}%
\begin{tabular}{|c|c|c|c|c|}
\hline 
\multirow{2}{*}{Width (nm)} & \multicolumn{4}{c}{Energy gaps (eV) for \textbf{3p} series of AGNRs}\tabularnewline
\cline{2-5} 
 & GW\citet{yang2} & DFT\citet{yang2} & \selectlanguage{english}%
$U=6.0$ eV\selectlanguage{american}
 & \selectlanguage{english}%
$U=8.0$ eV\selectlanguage{american}
\tabularnewline
\hline 
0.61 & 2.72 & 1.13 & 2.31 & 2.65\tabularnewline
\hline 
0.98 & 2.01 & 0.68 & 1.39 & 2.01 \tabularnewline
\hline 
1.35 & 1.68 & 0.55 & 1.17 & 1.63\tabularnewline
\end{tabular}

\begin{tabular}{|c|c|c|c|c|}
\hline 
\multirow{2}{*}{Width (nm)} & \multicolumn{4}{c}{Energy gaps (eV) for \textbf{3p+1} series of AGNRs}\tabularnewline
\cline{2-5} 
 & GW\citet{yang2} & DFT\citet{yang2} & \selectlanguage{english}%
$U=6.0$ eV\selectlanguage{american}
 & \selectlanguage{english}%
$U=8.0$ eV\selectlanguage{american}
\tabularnewline
\hline 
0.36893 & 5.5 & 2.50 & 3.29 & 3.72\tabularnewline
\hline 
0.73785 & 3.83 & 1.60 & 2.18 & 2.50\tabularnewline
\hline 
1.10678 & 2.83 & 1.13 & 1.64 & 1.90\tabularnewline
\hline 
1.47571 & 2.40 & 0.93 & 1.33 & 1.55\tabularnewline
\hline 
\end{tabular}

\begin{tabular}{|c|c|c|c|c|}
\hline 
\multirow{2}{*}{Width (nm)} & \multicolumn{4}{c}{Energy gaps (eV) for \textbf{3p+2} series of AGNRs}\tabularnewline
\cline{2-5} 
 & GW\citet{yang2} & DFT\citet{yang2} & \selectlanguage{english}%
$U=6.0$ eV\selectlanguage{american}
 & \selectlanguage{english}%
$U=8.0$ eV\selectlanguage{american}
\tabularnewline
\hline 
0.49190 & 1.71 & 0.47 & 0.41 & 0.67\tabularnewline
\hline 
0.86083  & 1.18 & 0.33 & 0.31 & 0.50\tabularnewline
\hline 
1.22976 & 0.94 & 0.22 & 0.24 & 0.40\tabularnewline
\hline 
1.59868 & 0.77 & 0.19 & 0.20 & 0.33 \tabularnewline
\hline 
\end{tabular}\label{tab:AGNR-gaps}\selectlanguage{english}
\end{table}

\begin{table}

\caption{Variation of ZGNR band gaps with the ribbon width, computed using
various values of screened Coulomb parameters, and the P-P-P-UHF method,
compared with the \emph{ab initio }DFT and GW results\citet{yang2}.
Band gaps are also presented at the edge of the Brillouin Zone ($k=\pi/a$),
which are more or less width independent. All the gaps correspond
to the spin-polarized edge states of ZGNRs.}

\selectlanguage{american}%
\begin{tabular}{|c|c|c|c|c|c|}
\hline 
\multirow{2}{*}{Width (nm)} & \multicolumn{5}{c}{Energy gaps (eV) for ZGNRs}\tabularnewline
\cline{2-6} 
 & DFT\citet{yang2} & GW\citet{yang2} & $U=4.5$ eV & $U=6$ eV & $U=8$ eV\tabularnewline
\hline 
1.136 & 0.35 & 1.36 & 1.34 & 1.91 & 3.04\tabularnewline
\hline 
1.562 & 0.33 & 1.23 & 1.14 & 1.61 & 2.64\tabularnewline
\hline 
1.988 & 0.25 & 1.04 & 1.00 & 1.40 & 2.35\tabularnewline
\hline 
2.410 & 0.23 & 0.95 & 0.88 & 1.22 & 2.13\tabularnewline
\hline 
\end{tabular}\label{tab-ZGNR-gaps}

\begin{tabular}{|c|c|c|c|c|}
\hline 
\multicolumn{5}{|c}{Band gap (eV) at the zone boundary ($k=\pi/a$) }\tabularnewline
\hline 
DFT\citet{yang2} & GW\citet{yang2} & $U=4.5$ eV & $U=6$ eV & $U=8$ eV\tabularnewline
\hline 
\hline 
$\backsim$0.45 & $\backsim$1.95 & $\backsim$1.94 & $\backsim$2.84 & $\backsim$4.40\tabularnewline
\hline 
\end{tabular}\selectlanguage{english}
\end{table}

An inspection of the two tables reveals the following trends: for
AGNRs, with $U=8.0$ eV excellent agreement is obtained between our
HF band gaps and the \emph{ab initio} GW band gaps\citet{yang2} of
the $N_{A}=3p$ family, however, for other families both with $U=6.0$
eV and $U=8.0$ eV our HF band gaps are smaller than the \emph{ab
initio} GW results but larger than the DFT results. For ZGNRs excellent
agreement is obtained between our P-P-P-UHF results and the GW results\citet{yang2}
for $U=4.5$ eV, while with all the larger values of $U$, our approach
overestimates the results as compared to the GW ones. Thus, we conclude
that no single value of Coulomb parameter $U$ provides good agreement
between our HF results and GW results for all classes of GNRs. Therefore,
in the absence of any reliable experimental data on the GNR band gaps,
we have decided to use the original screened parameters \textcolor{black}{of
Chandross and Mazumdar\citet{chandross}, with $U=8.0$ eV and $\kappa_{i,j}=2.0$
($i\neq j)$ and $\kappa_{i,i}=1$.}

\paragraph{AGNR Band Structure}

\begin{figure}
\includegraphics[width=8cm]{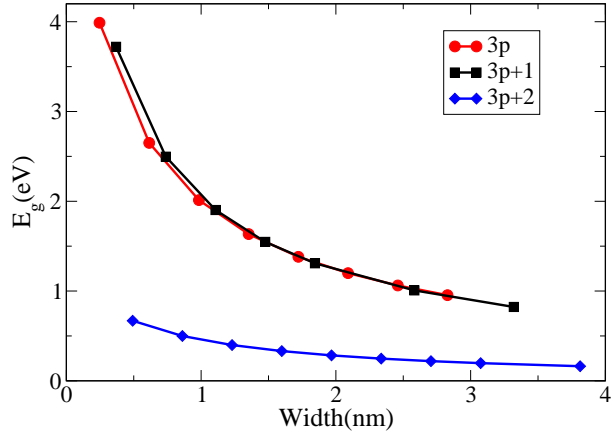}

\caption{(Color online) Variation of the band gaps of AGNRs-$N_{A}$, for $N_{A}=3p,\;3p+1,$
and $3p+2$, families, with respect to their width. All calculations
were performed with the original screened parameters\textcolor{black}{{}
($U=8.0$ eV and $\kappa_{i,j}=2.0$ ($i\neq j)$ and $\kappa_{i,i}=1$)}\citet{chandross}
in the P-P-P model. }

\label{fig:agnr-family}
\end{figure}

Before we discuss the band structure of AGNRs obtained using our approach,
we examine the variation of their band gaps with respect to the three
aforesaid families corresponding to the widths $N_{A}=3p,\;3p+1,$
and $3p+2$, where $p$ ($\geq0$) is an integer. This classification
is based upon the TB values of the band gaps which exhibit the relation
$E_{g}^{3p}\geq E_{g}^{3p+1}\geq E_{g}^{3p+2}(=0)$\citet{Son}. However,
\emph{ab initio} density-functional theory (DFT) calculations\citet{Son}
on these ribbons predicted a different relationship $E_{g}^{3p+1}\geq E_{g}^{3p}\geq E_{g}^{3p+2}(\neq0)$,
with the important result that even for $N_{A}=3p+2$, AGNRs exhibit
nonzero energy gaps, due to the fact that the bond lengths involving
the edge atoms are shorter than those in the interior. When the decrease
in the bond length is incorporated in the TB approach by increasing
the corresponding hopping, one also obtains finite gaps for $N_{A}=3p+2$
ribbons, although the relation $E_{g}^{3p}\geq E_{g}^{3p+1}\geq E_{g}^{3p+2}$
still holds. Here, we intend to investigate as to which of these relationships
holds when the AGNR band gaps are computed using our P-P-P-RHF approach.
In Fig. \ref{fig:agnr-family} we present the graph depicting the
variation of the band gaps of AGNRs for all the three families, with
respect to their width, computed using the screened Coulomb parameters
and our P-P-P-RHF method. From the figure it is obvious that while
the band gaps of $3p$ and $3p+1$ families are fairly close to each
other for a given width, yet the relation $E_{g}^{3p+1}\geq E_{g}^{3p}\geq E_{g}^{3p+2}(\neq0)$,
in agreement with the \emph{ab initio} DFT and GW results\citet{Son},
is found to hold for a fairly large range of width.

\begin{figure}
\includegraphics[width=10cm]{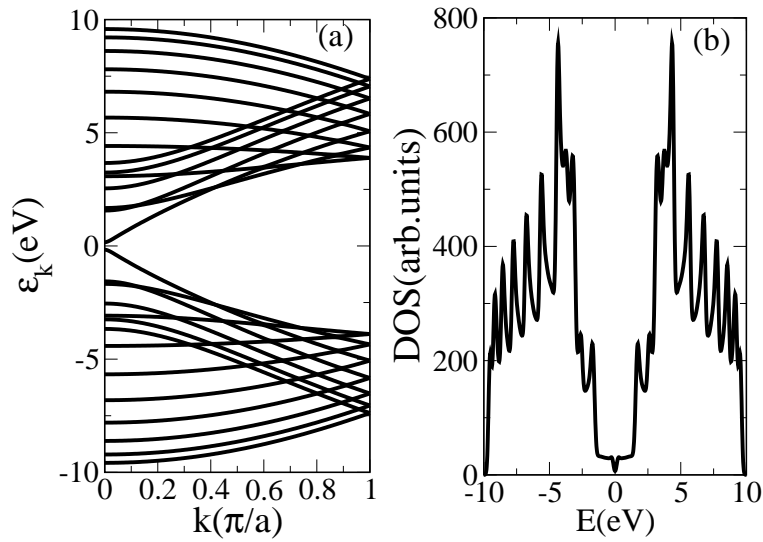}

\caption{(a) Band structure and (b) DOS of AGNR-14 obtained by P-P-P-RHF method,
obtained using the screened Coulomb parameters.}

\label{Flo:band-agnr}

\end{figure}

Next, in Fig. \ref{Flo:band-agnr}, we present the band structure
and the DOS of AGNR-14, belonging to the $3p+2$ family, computed
using our P-P-P-RHF method. It is evident from the figure that the
AGNR-14 is an insulating system with a small band gap of about 0.65
eV, located at the point $k=0$. The DOS presented in Fig. \ref{Flo:band-agnr}(b),
naturally vanishes in the region of the gap. Furthermore, it exhibits
several symmetrically placed peaks corresponding to the van Hove singularities. 

\paragraph{ZGNR Band Structure}

\begin{figure}
\includegraphics[width=10cm]{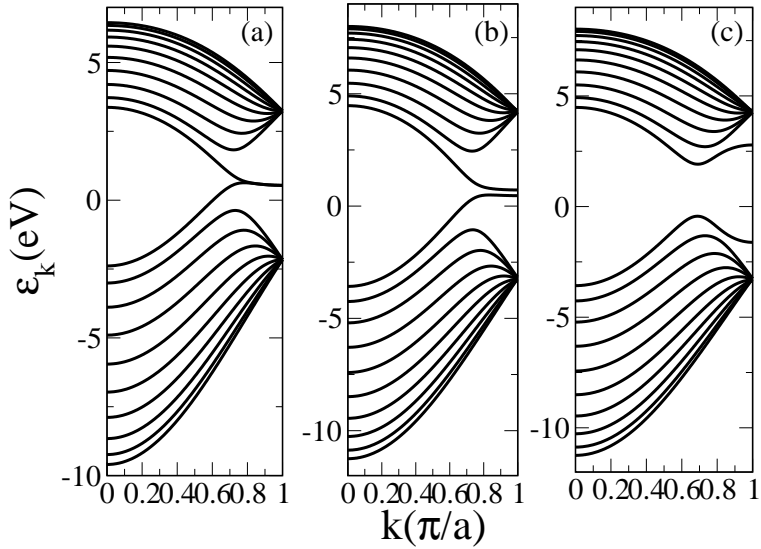}

\caption{Band structures of ZGNR-10 obtained by: (a) TB method (b) P-P-P-RHF
method, and (c) P-P-P-UHF method. }

\label{Flo:band-zgnr}
\end{figure}

\begin{figure}
\includegraphics[width=10cm]{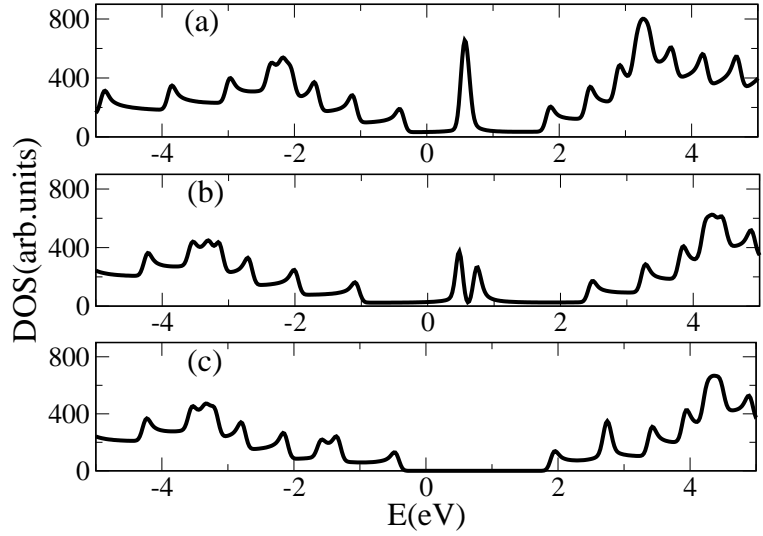}

\caption{The density of states of ZGNR-10 obtained by: (a) tight binding method,
(b) P-P-P-RHF method, and (c) P-P-P-UHF method. The Fermi energy $E_{F}$
is assumed to be at $E=0$. }

\label{Flo:dos-zgnr}
\end{figure}

The nature of the ground state of ZGNRs is quite interesting, with
several authors reporting the existence of a magnetically-ordered
ground state, with oppositely oriented spins localized on the zigzag
edges on the opposite sides of the ribbons\citet{fujita,ppp-nakada,okada},
a result verified also in several first principles DFT calculations\citet{Son,yang3}.
As reported in our earlier work, P-P-P-UHF calculations performed
using the present code also predict magnetized edge state as the ground
state of ZGNRs\citet{gundra1}. Here we elaborate the underlying physics
by performing TB, P-P-P-RHF and P-P-P-UHF calculations on ZGNR-10
using the present code, the results of which are presented in Figs.
\ref{Flo:band-zgnr} and \ref{Flo:dos-zgnr}. At the TB level ZGNR-10
(or a ZGNR of any other width) is obtained to be gapless as depicted
in Fig. \ref{Flo:band-zgnr}a, characterized by flat bands near $E_{F}$,
leading to a intense van Hove singularity at $E_{F}$ (\emph{cf}.
Fig.\ref{Flo:dos-zgnr}a). This suggests an instability in the system,
and, therefore, a possibility of a structural distortion through electron-phonon
coupling, or a magnetic ordering mediated by Coulomb interactions\citet{fujita}.
But, the band structure of ZGNR-10 obtained by the P-P-P-RHF method
(Fig.\ref{Flo:band-zgnr}b) is very similar to that obtained by the
TB method, except for a small band gap of about 0.25 eV, which is
an artifact of the RHF approach. Thus, the RHF method by its very
nature predicts a non-magnetic ground state, even though it takes
e-e interactions into account. However, once we perform spin-polarized
calculations using the UHF approach which is based upon separate mean-fields
for the up- and the down-spin electrons, we get the ground state exhibiting
edge magnetism, a significant band gap of 2.35 eV (\emph{cf}. Fig.\ref{Flo:band-zgnr}b),
and the total energy/cell of -55.532 eV which is  lower than -55.006
eV for the non-magnetic state obtained by the RHF calculations.

\begin{figure}

\includegraphics[angle=-90,width=10cm]{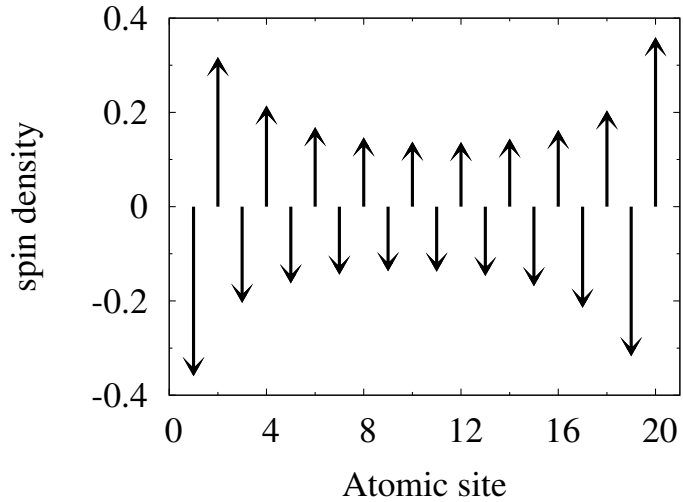}

\caption{Spin density of ZGNR-10, obtained using the P-P-P-UHF calculations,
plotted at different atomic sites of the unit cell across the width
of ribbon, starting from top. Anti-ferromagnetic alignment of spins
across the width is obvious.}

\label{fig:spin-den}
\end{figure}

The fact that ground state exhibits edge magnetism is obvious from
the spin-density plot for the ZGNR-10 obtained from the UHF calculations,
presented in Fig. \ref{fig:spin-den}. As far as the numerical aspects
of the UHF method are concerned, it is important to note that while
performing the UHF calculations the initial guesses for solutions
of the up- and down-spin Bloch orbitals must be different. Because,
for ZGNRs of any width, the number of electrons of spin up- and down-spin
are equal, as a result of which the UHF solutions converge to RHF
results, unless the initial guess for the two spin components is different. 

\begin{figure}
\includegraphics[width=10cm]{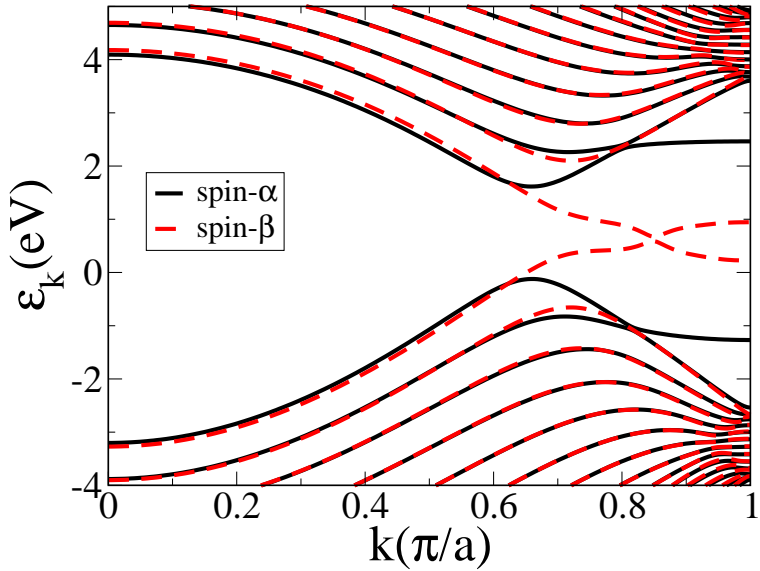}

\caption{(Color online) Band structure of ZGNR-14 in the presence of external
electric filed ($E_{y}$= 2 V/nm ), obtained by P-P-P-UHF method.
\textcolor{black}{The solid-black (dotted-red) lines represent the
bands of $\alpha(\beta)$-spin electrons.}}

\label{Flo:band-half-matallic}

\end{figure}

In the presence of a lateral electric field, ZGNRs exhibit half-metallic
behavior\citet{Son}, leading to their possible use in spintronics.
In Fig. \ref{Flo:band-half-matallic} we present the band structure
of the ZGNR-14 exposed to a field strength of 2 V/nm. The calculations
were performed using the P-P-P-UHF model with $U=8$. In the absence
of the field, as discussed above, the bands of the up- ($\alpha$)
and down-spin ($\beta$) electrons are degenerate with a band gap
of 1.96 eV . The degeneracy is lifted in the presence of electric
filed, the band gap for electrons of spin $\alpha$ is changes to
1.74 eV and the that for electrons of spin $\beta$ changes to 0.08
eV, indicating the half-metallic nature. \textcolor{black}{The band
gap and the critical field strength ($E_{y}^{c}$) to achieve the
half-metallicity for a given ZGNR decreases with the decrease in the
value of the Coulomb parameter $U$. For example, in case of ZGNR-14,
with $U=4.5$ eV, the band gap for electrons of spin $\beta$($\alpha$)
changes to 0.06 (0.76) eV, when the ribbon is subjected to a electric
filed strength of 1.0 V/nm, from the gap of 0.79 eV, in the absence
of the field. Furthermore, for a fixed value of $U$, $E_{y}^{c}$
decreases with increasing the width ($w$) of the ZGNR, due to the
decrease in the band gap and increase in the potential difference
($V=wE_{y}^{c}$) between the two edges of width $w$\citet{Son}}.

\subsection{Linear optical absorption spectrum}

\begin{figure}
\includegraphics[width=10cm]{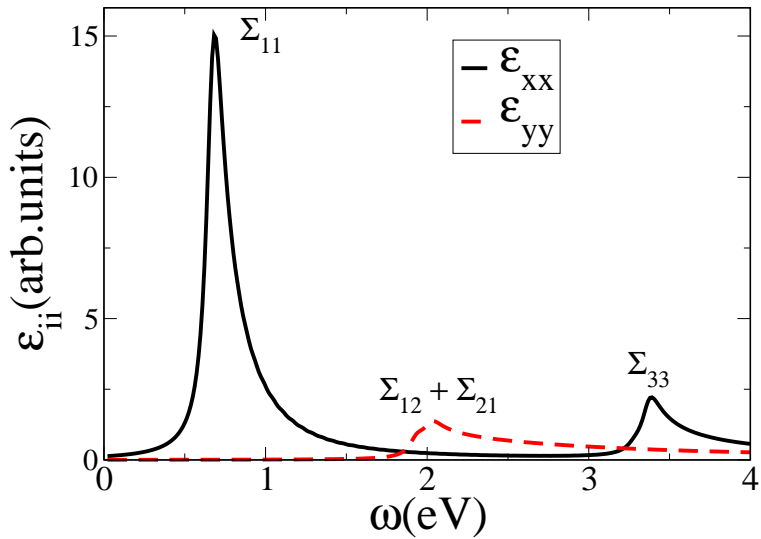}

\caption{(Color online) Optical absorption spectrum of AGNR-14 obtained by
P-P-P-RHF method. \textcolor{black}{The solid-black line represents
$\epsilon_{xx},$ while the dotted-red line represents $\epsilon_{yy}$.}
A line width of 0.05 eV was assumed.}

\label{Flo:opt-agnr}
\end{figure}

In our earlier work we argued that the polarization characteristics
of the optical absorption in GNRs is highly dependent on the nature
of its edges, and, thus, can be used to determine whether a ribbon
has zigzag or armchair edges\citet{gundra1}. In Fig. \ref{Flo:opt-agnr}
we present the optical absorption spectrum of the AGNR-14 obtained
using P-P-P-RHF method. If $\Sigma_{mn}$ denotes a peak in the spectrum
due to a transition from $m$-th valence band (counted from top) to
the $n$-th conduction band (counted from bottom), the peak of $\epsilon_{xx}(\omega)$
(black line) at 0.69 eV is $\Sigma_{11}$, at 3.40 eV is $\Sigma_{33}$.
The peaks of $\epsilon_{yy}(\omega)$ (red/dotted line) at 2.05 eV
correspond to $\Sigma_{12}$ and $\Sigma_{21}$. The noteworthy points
is that individual peaks in the spectrum correspond to either $x$-
or $y$-polarized photons, consistent with the $D_{2h}$ point group
of AGNRs. Furthermore, the $x-$ and $y-$polarized peaks are well
separated in energy. 

\begin{figure}
\includegraphics[width=10cm]{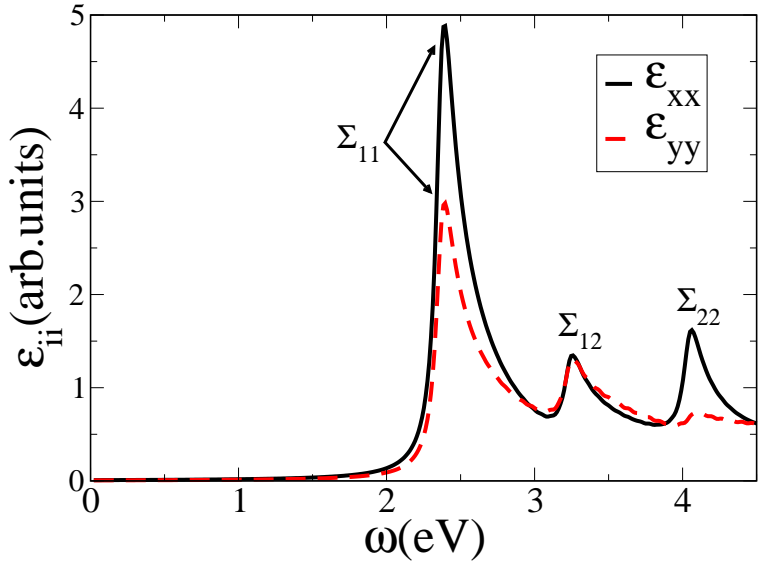}

\caption{(Color online) Optical absorption spectrum of ZGNR-10 obtained by
the P-P-P-UHF method. \textcolor{black}{The solid-black line represents
$\epsilon_{xx},$ while the dotted-red line represents $\epsilon_{yy}$.
}A line width of 0.05 eV was assumed.}

\label{Flo:opt-zgnr}

\end{figure}

In Fig. \ref{Flo:opt-zgnr} we present the optical absorption spectrum
of ZGNR-10 obtained using the P-P-P-UHF method. The peaks in $\epsilon_{xx}(\omega)$
are located at 2.41 eV ($\Sigma_{11}$), 3.20 eV ($\Sigma_{12}$),
and 4.07 ($\Sigma_{22}$), whereas the prominent peak of $\epsilon_{yy}(\omega)$
are at 2.41 eV ($\Sigma^{11}$) and 3.20 eV ($\Sigma_{12}$). Therefore,
in spite of the fact that the point group of ZGNRs is also $D_{2h}$,
yet unlike AGNRs, most of the prominent peaks of ZGNR-10 exhibit mixed
polarization characteristics. This is due to the fact that the edge-polarized
magnetic ground state of ZGNRs no longer exhibits $D_{2h}$ symmetry
because of the fact that the reflection symmetry about the $xz$-plane
is broken, thereby leading to mixed polarizations in the optical absorption.
Thus, by performing optical absorption experiments on oriented samples
of GNRs, one can predict whether a given ribbon is AGNR or ZGNR by
probing the polarization characteristics of the absorption peaks.

\subsection{Electro-absorption spectrum}

\begin{figure}
\includegraphics[width=10cm]{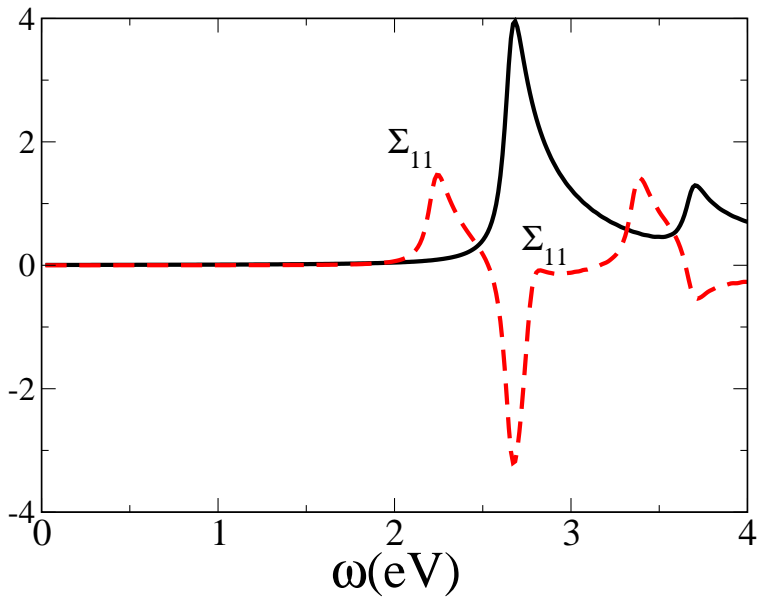}

\caption{(Color online) Linear absorption spectrum \textcolor{black}{(solid-black
line)}, and the electro-absorption spectrum \textcolor{black}{(dotted-red
line)} of ZGNR-8 obtained by the P-P-P-UHF method. }

\label{Flo:EA-zgnr}
\end{figure}

Electro-absorption (EA) spectroscopy, which consists of measuring
optical absorption in the presence of a static external electric field,
has been an important experimental tool for probing the electronic
structure and optical properties of conjugated polymers and other
materials\citet{EA-spec}. EA spectrum is defined as the difference
of the linear absorption spectra with, and without, an external static
electric field. In our earlier work we argued that the EA spectrum
can be used as a probe of both the electric-field driven half-metallicity,
as well as the edge magnetism of ZGNRs\citet{gundra1}. The essential
physics behind it is that in the presence of a lateral external electric
field, the band gap of a spin-polarized ZGNR for spin-up electrons
is different from those of down spins, leading to two split optical
transitions across the gap. We illustrate this in Fig. \ref{Flo:EA-zgnr}
which contains the EA spectrum \textcolor{black}{of ZGNR-8 in the
presence of a lateral external electric field of strength 1 V/nm}\textcolor{green}{,}
as well as the linear absorption without the field, for its spin-polarized
ground state, computed using the P-P-P-UHF approach. The tendency
towards half metallicity is apparent with the presence of two energetically
split peaks corresponding to two different $\Sigma_{11}$ transitions
among up and down-spin electrons. Noteworthy point is that the two
split peaks across the fundamental gaps will occur only if the ZGNR
has a edge-magnetized ground state, and not when the system has a
non-magnetic ground state. Therefore, if these split peaks predicted
in our work can be verified in the EA spectra of the ZGNRs, it will
provide an all-optical probe of determining the nature of the edges
in GNRs.

\section{Conclusions and future directions\label{sec:Conclusions}}

In this paper we have described our Fortran 90 program which solves
the HF equations for both the closed- and open-shell 1D-periodic $\pi$-conjugated
systems using the TB and P-P-P models. The present computer program
has been written in Fortran 90 language which allows dynamic allocation
of memory, thereby freeing the code from artificial limits related
to array sizes. To demonstrate the capabilities of our code, we presented
results of numerous test calculations on various systems including
organic polymers, as well as GNRs. We presented the results of total
energy calculations on the polymer t-PA and PPP, while the band structure
and the density of states of t-PA and various GNRs were reported.
Furthermore, we also explored the electric-field driven half metallicity
of ZGNRs, as also the optical absorption spectra of both AGNRs and
ZGNRs. We also reported calculations on the EA spectrum of ZGNRs,
and argued that it can be used to determine the nature of edge termination
in GNRs so as to differentiate between armchair and zigzag type edges. 

Having developed a HF mean-field code for the 1D periodic $\pi$-conjugated
system, our next aim is to include the electron-correlation effects
for them, particularly for the band structure, as well as to account
for the excitonic effects in optical absorption. Work along those
directions is underway in our group, and results will be communicated
in future publications.

\paragraph{Acknowledgment}

We thank the Department of Science and Technology (DST), Government
of India, for providing financial support for this work under Grant
No. SR/S2/CMP-13/2006. K. G is grateful to Dr. S. V. G. Menon (BARC)
for his continued support of this work.

\end{document}